# Structural-Functional Analysis of Engineered Protein-Nanoparticle Assemblies Using Graphene Microelectrodes

Jinglei Ping,[a] Katherine W. Pulsipher,[b] Ramya Vishnubhotla,[a] Jose A. Villegas,[b] Tacey L. Hicks,[b] Stephanie Honig,[b] Jeffery G. Saven,[b] Ivan J. Dmochowski,[b] A. T. Charlie Johnson[a]

The characterization of protein-nanoparticle assemblies in solution remains a challenge. We demonstrate a technique based on a graphene microelectrode for structural-functional analysis of model systems composed of nanoparticles enclosed in open-pore and closed-pore ferritin molecules. The method readily resolves the difference in accessibility of the enclosed nanoparticle for charge transfer and offers the prospect for quantitative analysis of pore-mediated transport shed light on the spatial orientation of the protein subunits on the nanoparticle surface, faster and with higher sensitivity than conventional catalysis methods.

## Introduction

The ability to attach functional biomolecules to nanoparticle surfaces has spurred development of nano-therapeutic,[1] diagnostic,[2] and biosensing[3,4] agents, as well as novel nano-structures[5] and devices.[6] Methods for controlling the number and orientation of oligonucleotides and peptides at nanoparticle surfaces have been established,[6] but it remains challenging to create nanoparticle-protein assemblies with native-like protein structure and function.[7,8] One emerging paradigm is a thermophilic ferritin protein[9,10] whose 24 self-assembling four-helix bundles maintain native stoichiometry and secondary structure when encapsulating a single 6-nm gold nanoparticle (AuNP).[11-13] However, the assembly configuration in solution remains unknown because conventional methods for characterizing protein structure, such as X-ray crystallography,[14] are not suitable for liquid-phase protein-nanoparticle conjugates.

Here, we demonstrate a non-perturbing method using a graphene microelectrode[15] for structural-functional analysis of an ordered AuNP-ferritin protein assembly that differs substantively from an unstructured protein corona. Charge flowing from the AuNP through ferritin pores transfers into the graphene microelectrode and is recorded by an electrometer. The measurements are consistent with a pore diameter of 4.5-nm, providing evidence that ferritin maintains native-like quaternary structure upon AuNP encapsulation. This work highlights the design and characterization of nanoparticle-protein assemblies with tunable ionic conductivity and chemical reactivity, and demonstrates a new tool for sensitively probing protein-nanomaterial interactions.

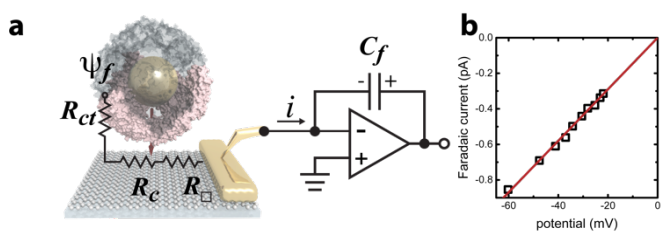

**Figure 1. a** Schematic of the setup for measuring spontaneous Faradaic charge transfer across a pore to a graphene microelectrode in buffer solution and circuit diagram. **b** Faradaic current as a function of electrostatic potential in the buffer solution above graphene. The red line is a linear fit to the data.

## Results and Discussion

Ferritin is a multimeric iron-storage protein comprising 24 protein subunits that self-assemble to form a hollow, ~8 nm inner diameter cage. The *Archaeoglobus fulgidus* ferritin (AfFtn) used here is a unique archaeal ferritin that forms a tetrahedral arrangement of its four-helix-bundle subunits, yielding four wide (4.5-nm), triangular pores spanning the 2-nm protein shell[16] (Fig. 1a). Stoichiometric addition of 6-nm gold nanoparticles (AuNPs) to disassembled apo-AfFtn induces AfFtn assembly around individual AuNPs capped with bis(*p*-sulfonatophenyl)phenylphosphine (BSSP),[11-13] while maintaining its native thermal stability, stoichiometry, ferroxidase activity, and secondary structure.[11] However, it is not understood whether AfFtn assembles in its native quaternary structure upon AuNP encapsulation, maintaining the large triangular pores, or whether subunits assemble in a more typical ferritin closed-pore conformation or adsorb in an unordered fashion.

A graphene microelectrode was used to quantify Faradaic current through a ferritin-AuNP assembly and thereby gain information about the arrangement of AfFtn subunits on the AuNP surface (i.e. differentiate between open- and closed-pore forms of the AfFtn shell). The experimental setup (Fig. 1a) consisted of a graphene-based microelectrode connected to the inverting input of an electrometer[15] (Keithley 6517a). The electrostatic potential above a protein assembly in fluid, $\psi_f$, drives a sub-picoampere Faradaic current, $i$, through the series resistance of the charge-transfer at the graphene-solution interface[15] ($R_{ct} \sim 100$ GΩ), the graphene sheet ($R_\square \sim 10^2 - 10^3$ Ω/$\square$), and the graphene-gold contact[17] ($R_c \sim 10$ Ω). Transferred charge accumulates on the feedback capacitor and is read out on the electrometer. Because there is no extrinsic bias voltage between the solution and the microelectrode, heat dissipation (aW μm$^{-2}$) and electrical perturbation (~pA) to the protein structure[18] are minimized. In a previous report[15] we documented the intrinsic low noise level for microelectrode measurements in an idealized buffer solution as well as excellent agreement between the data and theoretical models of the behaviour of the electric double layer above graphene.

The Faradaic current $i$ is proportional to the potential $\psi_f$: $i = \psi_f / R_{ct}$. We applied a phosphate buffer solution to the graphene microelectrode and measured the Faradaic current while the electrostatic potential above the graphene surface was tuned by varying the buffer ionic strength $c$ (Fig. 1b). The variation of $\psi_f$ with ionic strength was inferred from the Grahame equation for the electric double layer. The fit to the data corresponds to a constant charge-transfer resistance $R_{ct}$ = 69 ± 2 GΩ, and the fit passes through the origin as expected (0.6 ± 0.9 fA).

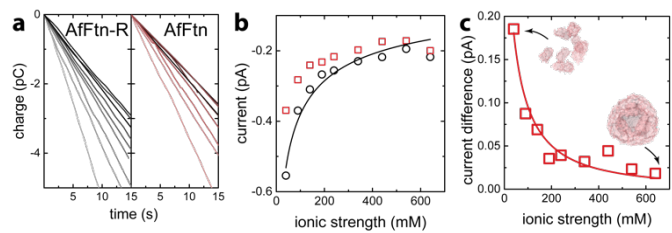

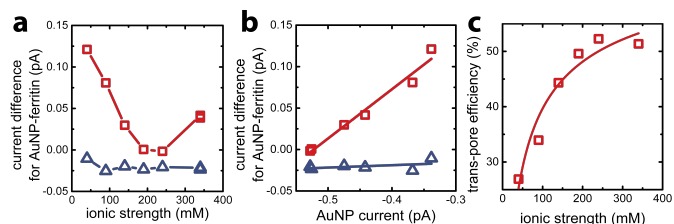

**Figure 2.** a Time trace of the charge transfer between a graphene microelectrode and mutant *A. fulgidus* E65R ferritin solution (AfFtn-R), and wild-type *A. fulgidus* ferritin solution (AfFtn). The ionic strength of the solution increases as the grey-level of the data increases. b Faradaic current for AfFtn-R (black circles) and AfFtn (red squares) based on the data in panel a. The black curve is a fit to the data for AfFtn-R using Supplemental Eqn. (1S). c Faradaic current difference for AfFtn compared to AfFtn-R; the red curve is a fit based on Supplemental Eqn. (2S).

The assembly state of AfFtn in solution is affected by ionic strength: it assembles into the native 24mer cage at high ionic strength and disassembles into dimers at low ionic strength.[16] For quantifying trans-pore current *via* the enclosed AuNP, the current baseline for the AuNP-ferritin system was determined by measuring the Faradaic current of an AfFtn mutant (E65R, termed AfFtn-R), which remains a 24mer even in low ionic strength solution. (See Supplementary Fig. S1.) The solution (200 μL; 20 nM) was applied to the microelectrode, and a sparse layer of non-specifically bound protein allowed to form and equilibrate (Supplementary Fig. S2). As the ionic strength of the solution was varied from 40 mM to 640 mM, a 15-sec time trace of Faradaic charge transfer (Fig. 2a) for AfFtn-R or AfFtn was used to extract the corresponding Faradaic current (Fig. 2b). Because of its excellent linearity, this 15-sec time trace is sufficient to determine the Faradaic current with high accuracy: indeed, for all figures in this report, the statistical errors associated with the electronic measurement are smaller than the size of the plotted points, and the observed scatter in the data is ascribed to experimental variation in the biofluid that is difficult to control. The solutions showed nearly identical Faradaic current at high ionic strength where both ferritins form stable 24mer assemblies, but the currents differed significantly at low ionic strength, where only AfFtn disassembles into dimers. The measured current for the AfFtn-R solution (black circles in Fig 2b) and the Faradaic current difference between the solutions of AfFtn and AfFtn-R (Fig. 2c) are well explained by models based on known properties of the electric double layer and AfFtn assembly. In particular, we infer a dissociation constant for AfFtn of 210 ± 60 mM, in agreement the value found from liquid chromatography measurements. (See Supplementary Section 2 for details.)

**Figure 3.** a Difference in Faradaic current for solutions of AuNP-enclosed in *A. fulgidus* mutant ferritin K150A/R151A (AuNP-AfFtn-AA, blue triangles) and AuNP-enclosed in wild-type ferritin (AuNP-AfFtn, red squares) compared to the baseline current set by a solution of E65R ferritin (AfFtn-R). Two data points, which almost overlap with each other, were tested at the concentration of 340 mM for both samples. b Faradaic current difference for AuNP-AfFtn-AA (blue triangles) and AuNP-AfFtn (red squares) as a function of the Faradaic current for AuNP. The lines are linear fits to the data. c Charge-transfer efficiency $\xi$ as a function of ionic strength fitted by the formula for the model based on electrical double layer.

To assess the configuration of AfFtn subunits on the surface of an AuNP, we measured the real-time Faradaic charge transfer for solutions of AuNPs ($I_{AuNP}$, see Supplementary Fig. S3), and of AuNP-ferritin assemblies based upon the wild-type ferritin AfFtn and a recently identified mutant, AfFtn-AA (K150A/R151A), which features an octahedral arrangement of its subunits with "closed" (< 1 nm) pores.[19] Representative data are shown in Supplementary Fig. S4. If AfFtn and AfFtn-AA maintain their native quaternary structure upon AuNP encapsulation, the AuNP surface should be less accessible for AfFtn-AA compared to AfFtn, and there should therefore be less charge transfer.

We used the Faradaic current of AfFtn-R as the baseline for assembled (24mer) ferritin, which leads to several strict requirements for accurate quantification of trans-pore current. First, ferritin must remain assembled with the AuNP enclosed. This is satisfied as 24mer assemblies of both AuNP-AfFtn and AuNP-AfFtn-AA are stable in the range of ionic strengths tested (40 – 340 mM).[11] Second, all AuNPs must be encapsulated by ferritin with no free AuNPs in solution. As shown in Supplementary Fig. S5 and S6, more than 99% of AuNPs were enclosed in a ferritin protein shell as confirmed by TEM and gel electrophoresis. We also verified that AuNPs were stable in the range of ionic strengths used without aggregation (Supplementary Fig. S7).

To quantify Faradaic current contributed by enclosed AuNPs, we calculated the difference between the current for AuNP-AfFtn (AuNP-AfFtn-AA), $I_{\text{AuNP-AfFtn}}$ ($I_{\text{AuNP-AfFtn-AA}}$) and the baseline ($I_{\text{AfFtn-R}}$): $\Delta I = I_{\text{AuNP-AfFtn}} - I_{\text{AfFtn-R}}$ ($I_{\text{AuNP-AfFtn-AA}} - I_{\text{AfFtn-R}}$), with results plotted in Fig. 3a. For AuNP-AfFtn, $\Delta I$ varied by ~0.12 pA through the range of ionic strength, with a minimum at ~240 mM. For AuNP-AfFtn-AA, $\Delta I$ was much smaller and essentially constant at -0.020 ± 0.005 pA. For AuNP-AfFtn, the plot of $\Delta I$ vs. $I_{\text{AuNP}}$ (Fig. 3b) followed a linear trend with slope $a$ = 0.59 ± 0.05, suggesting that the efficiency of Faradaic charge transfer via AuNPs enclosed in open-pore AfFtn is ~60% of that for bare AuNPs. In contrast, for AuNP-AfFtn-AA, we found a slope $a$ = 0.03 ± 0.03, suggesting that the ferritin closed pores completely suppress this charge transfer pathway.

This analysis suggests that the Faradaic current carried by the ferritin-AuNP system has two components: i) pore-mediated current via the AuNP and ii) current associated with the protein shell, quantified by $I_{\text{AfFtn-R}}$. We define the trans-pore efficiency $\xi(c) = |aI_{\text{AuNP}}(c)|/\left(|aI_{\text{AuNP}}(c)| + |I_{\text{AfFtn-R}}(c)|\right)$ to quantify the fraction of the total current carried by the enclosed AuNPs. The efficiency increases monotonically by ~100% as the ionic strength increases from 40 mM to 340 mM (Fig. 3c). In contrast to molecular diffusion through the pore, which is driven by a concentration gradient, the Faradaic current depends on the gradient of the electrostatic potential. Thus, negative charge at the edge of the AfFtn pores can suppress the (negative) Faradaic current via the enclosed AuNP over length scales given by the Debye screening length $\lambda_D[nm] = 0.304/\sqrt{c[M]}$. Thus we expect that the efficiency will be affected by ionic strength approximately as $\xi = A(4.5 - k\lambda_D[nm])^2$ where $A$ is a factor scaling area to efficiency, 4.5 nm is the pore diameter for AfFtn, and $k$ is ~1. The charge-transfer efficiency is well fit by this equation (Fig. 3c) with best fit value $k$ = 1.2 ± 0.1. This experiment demonstrates the capability of graphene microelectrode measurements to differentiate between open- and closed-pore structures in ferritin-nanoparticle assemblies, confirms the solvent accessibility of enclosed AuNPs, and provides strong evidence that the AfFtn pore maintains a native-like structure in the presence of the enclosed AuNP.

For confirmation and comparison, we used conventional catalysis methods to differentiate between wild-type AuNP-AfFtn and AuNP-AfFtn-AA: dehalogenation of a bisiodinated boron dipyrromethene derivative (I-BODIPY) and reduction of 4-nitrophenol. More AuNP surface area should be exposed in the AfFtn-containing sample compared to AfFtn-AA, and should therefore have greater AuNP catalytic activity. The reactions were monitored by different spectroscopic techniques: an increase in fluorescence at 507 nm was observed for the I-BODIPY dehalogenation reaction,[20] and a decrease in absorbance at 400 nm was observed for the 4-nitrophenol reduction.[21] The mechanism for AuNP-catalyzed dehalogenation of I-BODIPY is not well-understood but appears to be pseudo-zero order based on our data, similar to what was observed for dehalogenation of iodobenzene by AuNPs.[22] An induction period was observed in the 4-nitrophenol reduction, similar to polymer-coated AuNP systems.[23-26] This induction period has been attributed to diffusion of reagents to the AuNP surface and surface rearrangement of the AuNP before reaction can occur.[24] We expect similar effects to be in play for our AfFtn-coated AuNPs.

As shown in Fig. 4a, the rate of increase in the fluorescence intensity in the AuNP-AfFtn solution (0.0081 ± 0.0002 A.U./min) is approximately 4 times larger than the AuNP-AfFtn-AA solution (0.0019 ± 0.0002 A.U./min). For the 4-nitrophenol reduction, AuNP-AfFtn had roughly twice the catalytic rate constant, $k$ = (7.4 ± 0.7) x$10^{-3}$ s$^{-1}$ vs. (4.0 ± 0.3) x$10^{-3}$ s$^{-1}$ for AuNP-AfFtn-AA (Fig. 4b). Neither ferritin contributed to the catalytic activity; see Supplementary Fig. S8. For the catalytic assays, the difference in signal for AuNP-AfFtn versus AuNP-AfFtn-AA is only four-fold and two-fold for the I-BODIPY and 4-nitrophenol reactions, respectively. In contrast, the difference between AuNP-AfFtn and AuNP-AfFtn-AA for the microelectrode is nearly 20-fold (Fig. 3c). Thus, our methodology based on graphene microelectrode is comparatively rapid (seconds vs tens of minutes), has the potential for a quantitative estimate of the pore diameter through direct charge-transfer measurement through the protein shell, and could overcome limitations in sensitivity imposed by the AuNP catalytic reactions. FInally, the electrode-based method only requires relatively small amounts of sample solution (~ tens of μL) compared to the catalytic method, which also requires signficant amounts of additional reagents (I-BODIPY, 4-nitrophenol, and NaBH$_4$).

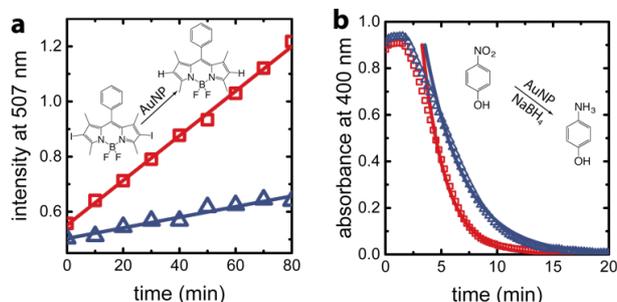

**Figure 4. a** Real-time fluorescence intensity of I-BODIPY dehalogenation catalyzed by AuNP-AfFtn-AA (blue triangles) and AuNP-AfFtn (red squares) solutions. For each measurement, 10 nM AuNP-AfFtn and 1 μM I-BODIPY were mixed in 50 mM HEPES, pH 7.0 (100 μL total volume). **b** Real-time UV-visible spectroscopy of reduction of 4-nitrophenol catalyzed by AuNP-AfFtn-AA and AuNP-AfFtn solutions. For each measurement, 5 nM AuNP-AfFtn and 50 μM 4-nitrophenol were mixed in a cuvette. Freshly prepared aqueous NaBH$_4$ was added to a final concentration of 2.5 mM and total sample volume of 1 mL. The solid curves are fits based on first-order kinetics. The corresponding catalytic reactions are shown in panels a-b.

## Conclusions

We have developed a graphene microelectrode device as a sensitive tool for structure-function analysis of AuNP-ferritin assemblies in solution. This all-electronic method has multiple advantages for identifying protein pores compared to conventional AuNP catalysis methods, and it has the potential to be developed into a direct measurement of the pore-mediated charge-transfer process. Our approach could provide a way to explore protein structure at nm-scale and, more broadly, to explore interactions of biomolecules with inorganic nanomaterials in complex biofluids--systems shown to offer

significant promise in bio-imaging, sensing,[4] catalysis and templated nanoparticle synthesis.[7]

## Methods

**Graphene growth.** Copper foil (99.8% purity) was loaded into a four-inch quartz tube furnace and annealed for 30 min at 1050 °C in ultra-high-purity (99.999%) hydrogen atmosphere (flow rate 80 sccm; pressure 850mT at the tube outlet) for removal of oxide residues. Graphene was deposited by low-pressure chemical vapor deposition (CVD) using methane as a precursor (flow rate 45 sccm) at 1050 °C for 60 min.

**Graphene device fabrication.** The graphene-copper growth substrate was coated with 500-nm layer of poly(methyl methacrylate) (PMMA, MicroChem), and the PMMA-graphene film was floated off the surface by immersion in 0.1 M NaOH solution with the graphene-copper growth substrate connected to the cathode of a power supply. The PMMA-graphene film was transferred onto a 300 nm $SiO_2$/Si wafer with an array of 5 nm/40 nm Cr/Au contact electrodes that were previously fabricated using photolithography. $Al_2O_3$ (5 nm) was deposited on the whole wafer as a sacrificial layer, and 50 μm × 100 μm graphene microelectrodes were defined by photolithography with photoresist PMGI (MicroChem) and S1813 (MICROPOSIT) and oxygen plasma etching. The $Al_2O_3$ layer on top of the microelectrode areas was removed by the basic photoresist developer MIF-319 (MICROPOSIT). After removal of the photoresist residues with 1165 (MICROPOSIT), another passivation layer of photoresist SU-8 (MicroChem) was applied to the device, and the wells exposing the microelectrodes were defined via photolithography.

**Computational Design of AfFtn-R.** The crystal structure of wild-type AfFtn (PDB 1SQ3) reveals a trimeric interface lined with negatively charged amino acids. We hypothesized that a high ionic strength solution allows for assembly by shielding the electrostatic repulsion between subunits at this interface. Therefore, a point mutation that inserts a positive charge along the interface, with potential for forming salt bridges to a neighboring subunit, could stabilize the 24mer cage at low ionic strengths.

We employed the statistical computational aided design strategy to guide the selection of stabilizing point mutations.[27-30] Calculations were carried out using atomic coordinates from chains G, H and J of AfFtn in PDB structure 1S3Q[2]. For each calculation, all amino acids were considered at the site selected, except for cysteine and proline. All other sites other than the site of interest were constrained to the wild-type identity and crystal structure conformation. Side-chain conformational states were taken from a rotamer library, and all possible conformations were considered.[31] Hydrogen atoms were placed according to the CHARMM19 topology files.[32] Energies were calculated using the CHARMM19 dihedral, van der Waals, and electrostatic terms were considered, with a non-bonded cut-off of 8 Å. Amino acid probability profiles were generated by summing the rotamer probabilities of each amino acid type.

Sites 34 and 65 of AtFtn are at the center of the carboxylate-rich pore. Analysis of these sites using the statistical computational design strategy recovered wild-type (glutamic acid, E) as the most probable amino acid at site 34. The most probable conformation possesses a favorable interaction with a neighboring positively charged lysine residue at site 39. This site was not selected for mutation. At site 65, the positively charged arginine (R) was the most probable residue. This site was chosen for mutation, and the arginine variant E65R was selected for expression.

**Ferritin mutagenesis.** The pAF0834 plasmid containing the AfFtn gene was provided by the laboratory of Dr. Eric Johnson at the California Institute of Technology. AfFtn-R was made by site-directed mutagenesis using the Stratagene QuikChange kit. The primers were obtained from Integrated DNA Technologies: sense (5'-3') GATTTCGTTTCCCGTCGCGGTGGCCGTG, antisense (5'-3') CACGGCCACCGCGACGGGAAACGAAATC. The mutated cDNA was transformed into XL1-Blue Supercompetent *E. coli* cells (Stratagene) according to the manufacturer's protocol. The plasmid was isolated using a QIAprep Spin Miniprep kit (Qiagen). All sequencing was performed by the University of Pennsylvania DNA Sequencing Facility. The AfFtn-AA plasmid was purchased from DNA2.0 and transformed the same as AfFtn-R.

**Ferritin expression/purification.** Production and purification of AfFtn and mutants was performed as previously published,[13] with some modifications. The plasmid was transformed into BL21-Codon Plus(DE3)-RP competent *E. coli* cells (Stratagene) in TB medium (1 L containing 100 mg ampicillin, 35 mg chloramphenicol) at 37 °C with shaking at 225 rpm until $OD_{600}$ ~0.8 was reached. Expression was induced with 1 mM isopropyl β-D-1-thiogalactopyranoside (IPTG, Lab Scientific) and incubation at 37 °C was continued for 4 h. Cells were centrifuged and stored at -20 °C, followed by resuspension in buffer (20 mM phosphate, 20 mM NaCl, pH 7.6) with an EDTA-free protease inhibitor cocktail tablet (ThermoFisher Scientific). Cell lysis was performed by treatment with lysozyme (~1 mg/mL final concentration) and sonication (amplitude of 30, 1 s on, 1 s off, 10 min total processing time). Cellular debris was removed by centrifugation (6 krpm, 30 min, 4 °C), and the supernatant was treated with nuclease (Pierce universal nuclease, ThermoFisher Scientific) after addition of $MgCl_2$ to a final concentration of 2 mM for 15 min at room temperature. The solution was heat shocked to remove most endogenous *E. coli* proteins (10 min at 80 °C). After pelleting the precipitated *E. coli* proteins by centrifugation (9 krpm, 60 min, 4 °C), the supernatant was buffer exchanged to ensure complete ferritin assembly (2.5 M NaCl, 2 mM EDTA, 20 mM phosphate, pH 7.6), and purified further using size exclusion chromatography (HiLoad 16/60 column, GE Healthcare). The purity of the protein was determined to be >95% by denaturing PAGE gel (4-15% Tris-HCl, Mini-Protean TGX gel), as seen in Supplementary Fig. S9a. Protein concentration was determined using the Bio-Rad Protein Assay (based on the Bradford method), using bovine gamma globulin as the standard. Proteins were also characterized by MALDI-TOF-MS, TEM, and DLS (see Fig. S9b, Fig. S9c and Table S1 in Supplementary Information). Protein stock solutions were 0.22 µm filtered and stored at 4 °C until use in experiments. Multiple stock solutions of ferritins were used for experiments to ensure reproducibility.

**AfFtn solution and AfFtn-R solution preparation.** Protein samples were prepared at 10 µM concentration in phosphate buffer (20 mM phosphate, pH 7.6), using NaCl to vary ionic strength (40, 90, 140, 190, 240, 340, 440, 540, 640 mM). To ensure accurate ionic strengths, samples were buffer exchanged on a Zebaspin column (ThermoFisher Scientific) equilibrated with the appropriate buffer. Samples were incubated overnight at 4 °C to allow for equilibration.

**AuNP-AfFtn solution and AuNP-AfFtn-AA solution preparation.** Citrate-capped 6-nm AuNPs were purchased from TedPella. The citrate was exchanged for bis(p-sulfonatophenyl)phenylphosphine (BSPP, Strem Chemicals) as described previously.[33] For device measurements, 1 mL samples were prepared at 0.3 mg/mL protein, 0.6 µM 6-nm AuNP-BSPP in 20 mM phosphate pH 7.6 and equilibrated at room temperature for 48 h with gentle agitation to ensure encapsulation. Protein NP samples were buffer-exchanged into various ionic strengths (40, 90, 140, 190, 240, 340 mM) using 10DG columns (Bio-Rad) equilibrated with the appropriate phosphate buffer. The 10DG column also helped ensure that only encapsulated AuNPs remained in the samples, as confirmed by TEM and native gel electrophoresis (see Supplementary Fig. S5 and S6). The first two fractions were combined, and AuNP concentration was verified by UV-vis spectroscopy. Because bare AuNPs cannot elute on a 10DG column, buffer exchange for the AuNP samples without protein was done using Zebaspin columns equilibrated at the same ionic strengths. All samples were diluted to 2 mL to match the lowest concentration sample (20 nM). All samples were measured on the same device on the same day they were prepared, to minimize bulk AuNP aggregation.

**Preparation of I-BODIPY.** I-BODIPY was prepared following the method of Zuber et al.[34] A dark red solid product was obtained with a mass of 31.8 mg (69.7% yield). $^1$H NMR (CD$_2$Cl$_2$): 7.54 (3H, m), 7.29-7.28 (2H, m), 2.62 (6H, s), 1.40 (6 H, s). Mass was verified using MALDI-TOF-MS with α-cyano-4-hydroxycinnamic acid (CHCA) as matrix. For characterization data, see Supplementary Fig. S10.

**Fluorescence measurements.** For the AuNP-catalyzed dehalogenation reaction, 10 nM AuNP-AfFtn and 1 µM I-BODIPY were mixed in 50 mM HEPES buffer (pH 7.0). Steady-state fluorescence was monitored using a Varian Cary Eclipse fluorimeter, with PMT detector voltage at 800 V, excitation wavelength of 465 nm, and temperature of 25 °C.

**4-Nitrophenol reduction.** A solution of 5 nM AuNP-AfFtn and 50 µM 4-nitrophenol (Fluka) was mixed in a cuvette. Freshly prepared aqueous NaBH$_4$ (Fluka) was added to a final concentration of 2.5 mM and total sample volume of 1 mL. Absorbance at 200—1100 nm was measured every 15 s at 25 °C using an Agilent 8453 UV-visible spectrometer. To determine the rate constant $k$, the data were fit to a first-order reaction, after subtracting the induction time (197 s):

$$\text{Abs} = \varepsilon[A]_0 e^{-kt}$$

where Abs is the measured absorbance, $\varepsilon$ the extinction coefficient of 4-nitrophenol at 400 nm (18000 M$^{-1}$ cm$^{-1}$), $[A]_0$ the initial concentration of 4-nitrophenol (50 µM), and $t$ the time.

## Acknowledgements

J.P. and A.T.C.J. acknowledge the support from the Defense Advanced Research Projects Agency (DARPA) and the U.S. Army Research Office under grant number W911NF1010093. K.W.P. and I.J.D. acknowledge support from the NSF (PD 09-6885). J.G.S. and I.J.D. acknowledge support from NSF CHE-1508318. J.G.S. acknowledges additional support from the Penn Laboratory for Research on the Structure of Matter (NSF DMR-1120901). This work used the Extreme Science and Engineering Discovery Environment (XSEDE), which is supported by National Science Foundation grant number ACI-1053575, under grant number TG-CHE110041. MALDI-TOF-MS was purchased with NSF grant CHE-0820996. T.L.H. was supported by the REU program of the LRSM (Penn MRSEC) through NSF grant DMR-1062638 Penn-REU summer fellowship, NSF DMR 11-20901.

The authors thank the laboratory of Dr. Eric Johnson at the California Institute of Technology for providing the pAF0834 plasmid.

## References

1. E. Fantechi, C. Innocenti, M. Zanardelli, M. Fittipaldi, E. Falvo, M. Carbo, V. Shullani, L. D. C. Mannelli, C. Ghelardini, A. M. Ferretti, A. Ponti, C. Sangregorio and P. Ceci, *ACS Nano*, 2014, **8**, 4705-4719.
2. S. Kumar, J. Aaron and K. Sokolov, *Nature Protocols*, 2008, **3**, 314-320.
3. I. L. Medintz, H. T. Uyeda, E. R. Goldman and H. Mattoussi, *Nature Materials*, 2005, **4**, 435-446.
4. S. Rana, N. D. B. Le, R. Mout, K. Saha, G. Y. Tonga, R. E. S. B. Bain, O. R. Miranda, C. M. Rotello and V. M. Rotello, *Nature Nanotechnology*, 2015, **10**, 65-69.
5. Y. Suzuki, G. Cardone, D. Restrepo, P. D. Zavattieri, T. S. Baker and F. A. Tezcan, *Nature*, 2016, **533**, 369-373.
6. A. E. Nel, L. Madler, D. Velegol, T. Xia, E. M. V. Hoek, P. Somasundaran, F. Klaessig, V. Castranova and M. Thompson, *Nature Materials*, 2009, **8**, 543-556.
7. M. Lundqvist, J. Stigler, G. Elia, I. Lynch, T. Cedervall and K. A. Dawson, *Proceedings of the National Academy of Sciences of the United States of America*, 2008, **105**, 14265-14270.
8. B. Maity, S. Abe and T. Ueno, *Nature Communications*, 2017, **8**, 14820.
9. E. C. Theil, T. Tosha and R. K. Behera, *Accounts of chemical research*, 2016, **49**, 784-791.
10. M. Uchida, S. Kang, C. Reichhardt, K. Harlen and T. Douglas, *Biochimica et Biophysica Acta*, 2010, **1800**, 834-845.

11. J. C. Cheung-Lau, D. Liu, K. W. Pulsipher, W. Liu and I. J. Dmochowski, *J. Inorg. Biochem.*, 2014, **130**, 59-68.
12. J. Swift, C. A. Butts, J. Cheung-Lau, V. Yerubandi and I. J. Dmochowski, *Langmuir*, 2009, **25**, 5219-5225.
13. K. W. Pulsipher and I. J. Dmochowski, in *Methods in Molecular Biology*, ed. B. P. Orner, Springer New York, New York, NY, 2015, vol. 1252, pp. 27-37.
14. M. Kim, Y. Rho, K. S. Jin, B. Ahn, S. Jung, H. Kim and M. Ree, *Biomacromolecules*, 2011, **12**, 1629-1640.
15. J. Ping and A. T. C. Johnson, *Applied Physics Letters*, 2016, **109**, 013103.
16. E. Johnson, D. Cascio, M. R. Sawaya, M. Gingery and I. Schroder, *Structure*, 2005, **13**, 637-648.
17. Z. J. Qi, J. A. Rodriguez-Manzo, A. R. Botello-Mendez, S. J. Hong, E. A. Stach, Y. W. Park, J.-C. Charlier, M. Drndic and A. T. C. Johnson, *Nano Lett.*, 2014, **14**, 4238-4244.
18. Y. Luo, C. E. Ergenekan, J. T. Fischer, M.-L. Tan and T. Ichiye, *Biophys. J.*, 2010, **98**, 560-568.
19. B. Sana, E. Johnson, P. Le Magueres, A. Criswell, D. Cascio and S. Lim, *J. Biol. Chem.*, 2013, **288**, 32663-32672.
20. J. Park, S. Choi, T.-I. Kim and Y. Kim, *Analyst*, 2012, **137**, 4411-4414.
21. K. Esumi, K. Miyamoto and T. Yoshimura, *J. Colloid Interface Sci.*, 2002, **254**, 402-405.
22. A. Corma, R. Juarez, M. Boronat, F. Sanchez, M. Iglesias and H. Garcia, *Chem Commun (Camb)*, 2011, **47**, 1446-1448.
23. S. Gu, S. Wunder, Y. Lu, M. Ballauff, R. Fenger, K. Rademann, B. Jaquet and A. Zaccone, *The Journal of Physical Chemistry C*, 2014, **118**, 18618-18625.
24. S. Wunder, Y. Lu, M. Albrecht and M. Ballauff, *ACS Catalysis*, 2011, **1**, 908-916.
25. K. Kuroda, T. Ishida and M. Haruta, *Journal of Molecular Catalysis A: Chemical*, 2009, **298**, 7-11.
26. S. Wunder, F. Polzer, Y. Lu, Y. Mei and M. Ballauff, *Journal of Physical Chemistry C*, 2010, **114**, 8814-8820.
27. H. Kono and J. G. Saven, *J. Mol. Biol.*, 2001, **306**, 607-628.
28. J. R. Calhoun, H. Kono, S. Lahr, W. Wang, W. F. DeGrado and J. G. Saven, *J. Mol. Biol.*, 2003, **334**, 1101-1115.
29. G. M. Bender, A. Lehmann, H. Zou, H. Cheng, H. C. Fry, D. Engel, M. J. Therien, J. K. Blasie, H. Roder, J. G. Saven and W. F. DeGrado, *J. Am. Chem. Soc.*, 2007, **129**, 10732-10740.
30. C. A. Butts, J. Swift, S. G. Kang, L. Di Costanzo, D. W. Christianson, J. G. Saven and I. J. Dmochowski, *Biochemistry*, 2008, **47**, 12729-12739.
31. R. L. Dunbrack, Jr., *Curr. Opin. Struct. Biol.*, 2002, **12**, 431-440.
32. A. D. MacKerell, D. Bashford, M. Bellott, R. L. Dunbrack, J. D. Evanseck, M. J. Field, S. Fischer, J. Gao, H. Guo, S. Ha, D. Joseph-McCarthy, L. Kuchnir, K. Kuczera, F. T. Lau, C. Mattos, S. Michnick, T. Ngo, D. T. Nguyen, B. Prodhom, W. E. Reiher, B. Roux, M. Schlenkrich, J. C. Smith, R. Stote, J. Straub, M. Watanabe, J. Wiorkiewicz-Kuczera, D. Yin and M. Karplus, *J. Phys. Chem. B*, 1998, **102**, 3586-3616.
33. A. Marchetti, M. S. Parker, L. P. Moccia, E. O. Lin, A. L. Arrieta, F. Ribalet, M. E. P. Murphy, M. T. Maldonado and E. V. Armbrust, *Nature*, 2009, **457**, 467-470.
34. A. Zuber, M. Purdey, E. Schartner, C. Forbes, B. van der Hoek, D. Giles, A. Abell, T. Monro and H. Ebendorff-Heidepriem, *Sens. Actuators, B*, 2016, **227**, 117-127.